# A Mixed Methods Systematic Analysis of Issues and Factors Influencing Organizational Cloud Computing Adoption and Usage in the Public Sector: Initial Findings

Mark Theby

Capitol Technology University, Laurel, United States

*Abstract:* Cloud computing has been shown to be an essential enabling technology for public sector organizations (PSOs) and offers numerous potential benefits, including reduced information technology infrastructure costs, increased innovation potential, and improved resource resilience and scalability. Despite governments' intensifying efforts to realize the benefits of this technology, cloud computing adoption and usage proves to be challenging, posing a variety of organizational and operational issues for PSOs. This systematic analysis constitutes the initial phase of a larger research effort that involves forthcoming case studies of specific public sector cloud stakeholders; it aims to identify and synthesize the available knowledge on organizational cloud computing adoption and utilization in the public sector to provide public sector decision makers and stakeholders with reliable, evidence-based, actionable insights that inform and improve public sector IT practice and policy.

## I. INTRODUCTION

### A. Background

The concept of cloud computing alludes to the on-demand, convenient, and ubiquitous delivery of a wide range of pooled and configurable computing resources—such as applications, servers, storage, and development tools—via a computer network and has received increasing attention from a variety of organizations [1]. The cloud computing technology model has been widely embraced by the private sector, where it is providing extensive benefits such as reduced IT costs, greater agility and time-to-value, and improved scalability compared to legacy approaches for computing resource delivery [2], but has experienced a less robust adoption in the public sector [3].

For public sector organizations, the migration of IT resources and processes to the cloud can confer distinct benefits. Cloud computing can lower the upfront costs traditionally associated with resource-intensive computing and provides rapid access to hardware resources without requiring extensive capital investments [4]. Cloud computing can also enhance agility by enabling organizations to dynamically up- or down-scale their services dependent on operational requirements, making it feasible to meet sudden peak requirements without having to maintain costly slack resources [5]. By enabling public sector organizations to access state-of-the-art IT services at the same speed as the commercial world, and by supporting the rapid prototyping and development of tools and services, a transition to the cloud can also foster innovation [6]. Finally, as cloud users and cloud applications inherit accreditation controls from their central cloud platforms, the adoption of cloud computing can also increase the security of IT assets [7].

The distinct, potential benefits of cloud computing in the public sector only gained further importance during the COVID-19 pandemic [8]. Experiencing extraordinary challenges, governments and public sector agencies were forced to mount a rapid response to ensure continuity of operations, protect citizens, and address the economic and social fallout of the pandemic. Public sector work arrangements shifted, with employees being thrusted into remote work environments in record numbers, government services flexed radically to meet emergency demands, and critical new services were implemented to address essential government operations and policymaking needs. For government organizations that had cloud capabilities available, pivotal cloud advantages such as agility, scalability, and resiliency provided the foundation for a successful crisis response; for public sector organizations that lagged behind in cloud adoption, the crisis has further strengthened the case for expanded cloud computing [9]. Despite the maturity of cloud computing and its extensive utilization in the private sector, adoption and use in the public sector still lags in many countries [10].

*B. Rationale*

Cloud computing has been shown to be an essential enabling technology for public sector organizations (PSOs) and offers numerous potential benefits, including reduced information technology infrastructure costs, increased innovation potential, and improved resource resilience and scalability [11]. Despite governments' intensifying efforts to implement the technology and realize its benefits, cloud computing adoption and usage remains challenging, posing a variety of organizational and operational issues for PSOs. To provide public sector IT decision makers and practitioners with evidence-based, actionable information that informs and improves IT practice and policy, it is imperative to analyze and synthesize the available, state of the art scientific knowledge on cloud computing adoption and utilization in public sector settings.

Systematic and comprehensive efforts to analytically address public sector cloud computing adoption and utilization are limited in quantity and scope; currently, the extant literature only offers four such studies. One review from 2015 focuses on cloud adoption factors and assesses a pool of twelve primary studies that originated between 2010 and 2015 [12]. Despite a title that suggests an analysis of factors influencing the adoption of cloud computing for government implementation, the study heavily draws on private sector primary data for its analysis, raising significant questions about the applicability and value of the resulting findings for public sector settings [12]. Two research studies, one in 2019 and one in 2021, investigate the benefits and risks of cloud computing adoption in the public sector and cover fifty-one primary studies between 2011 and 2018 and twenty-two primary studies between 2015 and 2020, respectively [13], [14]. While these investigations of the benefits and risks of cloud adoption reveal important and relevant themes for public sector cloud practitioners and decision makers, the adoption and usage phenomenon of public sector cloud computing is considerably broader and deserves a holistic assessment.

Finally, a recent study from 2022 assesses twenty-eight primary studies that were conducted between 2015 and 2020 and identifies benefits, methodologies, and factors that affect cloud computing adoption in the public sector [15]. The study makes an insightful contribution that advances the mapping of the public sector research landscape and identifies research gaps, but its' discussion of the factors that affect cloud computing use and adoption is minimal and does not provide the level of detail necessary to inform IT practice and policy in PSOs. Additionally, [15] does not employ a rigorous critical appraisal process that allows for an assessment of methodological quality of the underlying data, raising questions whether its findings are empirically supported [16].

*C. Objective*

This mixed methods systematic analysis constitutes the initial phase of a larger research effort that involves forthcoming case studies of specific public sector cloud stakeholders; it aims to identify and synthesize the issues and factors that affect organizational cloud computing adoption and usage in the public sector. This study generates knowledge of the salient themes that have contributed to making cloud computing adoption and usage challenging for PSOs and provides a foundation for prioritizing efforts to improve the usage and adoption of cloud computing in public sector settings. In order to identify relevant research, the following top-level research question drives the inquiry in this study:

RQ: What are the major themes, i.e. issues and factors, in the organizational adoption and use of cloud computing in public sector organizations?

## II.   METHODS

*A. Protocol and Registration*

This study employs the Mixed Methods Systematic Review (MMSR) methodology [17] promulgated by the Joanna Briggs Institute (JBI) and follows a convergent integrated approach of synthesis and integration that acknowledges the methodological diversity of the underlying research domain. In line with JBI procedures, a research protocol was registered and published with the Center of Open Science (COS) prior to commencing the analysis (https://osf.io/hx38d/). The study follows Preferred Reporting Items for Systematic Reviews and Meta-Analyses (PRISMA) guidelines [18] and reports findings accordingly.

*B. Eligibility Criteria*

Only peer-reviewed scientific papers meeting all of the following criteria were included: (1) papers addressing the issues (qualitative) or factors (quantitative) that influence the organizational adoption or use of cloud computing in a public sector context; (2) published in the English language; (3) published before 1 June 2022; (4) utilizing any research design (i.e., qualitative, quantitative, multi-method or mixed methods). Research that did not exclusively report on public sector themes (i.e., included non-public sector organizations in the analysis) was excluded from this study to preclude the aggregation of inapplicable private sector issues or factors into this context-specific analysis. Since this study aims to explore primary data, editorial comments and systematic literature reviews were also excluded.

*C. Information Sources and Search*

ACM Digital Library, IEEE Xplore, and Scopus comprise the most-widely referenced and indexed peer-reviewed articles on information technology adoption and use and were systematically searched in July of 2022. The database search strategy is depicted in Table 1.

### TABLE 1: DATABASE SEARCH STRATEGY

| Information Source | Search Query |
|---|---|
| **IEEE Xplore** https://ieeexplore.ieee.org | ("Document Title":cloud AND "Document Title":government) OR ("Document Title":cloud AND "Document Title":"public sector") OR ("Document Title":cloud AND "Document Title":"public service") <br><br>Filters: Conferences   Journals |
| **ACM Digital Library** https://dl.acm.org/ | [Title: cloud] AND [[Title: "public sector"] OR [Title: "public service"] OR [Title: government]] |
| **Scopus** | TITLE ( cloud ) AND TITLE ( government OR "public sect*" OR "public serv*" ) AND ( LIMIT-TO ( DOCTYPE , "cp" ) OR LIMIT-TO ( DOCTYPE , "ar" ) OR LIMIT-TO ( DOCTYPE , "ch" ) ) |

*D. Study Selection and Data Collection*

The citations identified by the database searches were collated and uploaded into the EndNote reference management tool; duplicates were then removed. Titles and abstracts of the remaining citations were screened and assessed against the eligibility criteria outlined above; non-relevant studies were excluded based on this initial screening. The remaining citations were then screened and assessed a second time, this time based on a detailed review of the full-text of the items. At this stage, studies that did not meet all eligibility criteria were excluded and the reasons that lead to the exclusion were documented. Quantitative and qualitative data were then extracted from the final corpus of included studies.

*E. Data Transformation*

Following the guidelines of the JBI Manual for Evidence Synthesis [17], findings were extracted from the eligible studies and then thematic synthesized. Qualitative findings, including those obtained based on mixed methods research, were extracted as reflected in the primary level research. Quantitative findings, including those obtained from the quantitative section of mixed methods research studies, were converted into qualitized data, i.e. transformed into textual description and narrative interpretation of the quantitative studies.

*F. Risk of Bias in Individual Studies*

All eligible studies were critically appraised for methodological quality by utilizing the Mixed Methods Appraisal Tool (MMAT) [19], a widely-used critical appraisal tool designed for mixed methods research studies that allows for the assessment of a variety of study designs that include qualitative, quantitative, and mixed methods approaches [20]. All eligible studies met the methodological quality criteria: they possessed affirmative responses to the two baseline screening questions of the MMAT and rated four or more affirmative responses to the five study-design-focused questions of the tool [20].

*G. Qualitative Synthesis*

This study uses the thematic synthesis approach, a well-established methodology for data synthesis that is well-suited for the identification of themes and the analysis of empirical data that includes both quantitative and qualitative elements [16]. Thematic synthesis consists of three distinct stages: line-by-line coding, the creation of descriptive themes, and the development of analytical themes [21]. The approach, which initially remains close to the original findings of the primary studies in stages one and two but then gains analytical distance in stage three by going beyond the initial findings and producing new knowledge, balances descriptive and interpretive processes and allows for reciprocal translations that can create new constructs, explanations, or theories [16], [22].

The data set was inductively coded line-by-line using MAXQDA software (first stage) and then the codes were organized into descriptive themes, examining their similarities and differences (second stage). Finally, analytical themes were developed that allowed for the clustering of the descriptive themes to generate additional knowledge in relation to the objectives of this study (third stage).

### III.   RESULTS

*A. Study Selection*

The PRISMA 2020 Flow Diagram below (Fig. 1) outlines the studies that were identified, screened, and included in this mixed methods systematic analysis.

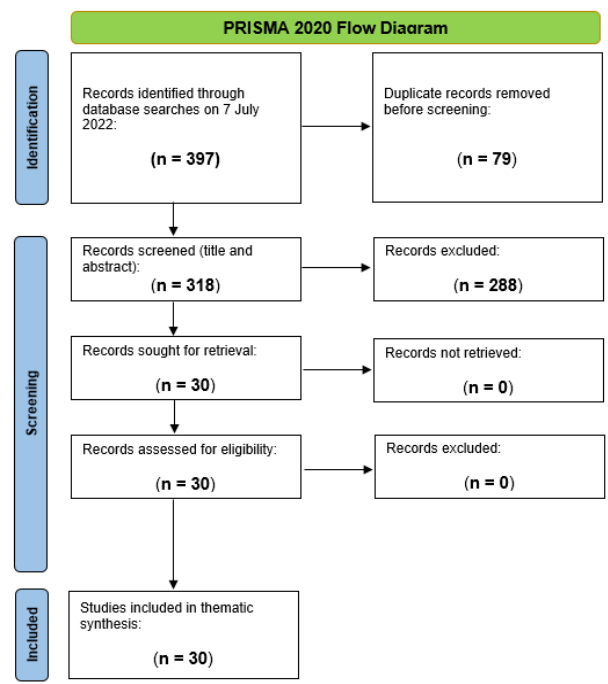

**Figure 1: PRISMA Flow Diagram [23]**

*B. Description of Studies included in this Analysis*

The main characteristics of the thirty studies included in this analysis are shown in Table 2. The studies reflect a geographically, theoretically, and methodologically diverse body of research that—despite its nascent nature—holds notable lessons for public sector decision makers and cloud practitioners.

In terms of methodology, thirteen of the studies utilize quantitative approaches, eleven studies follow qualitative strategies, and six studies apply mixed methods designs. In terms of theoretical foundations, eleven studies in the research corpus rely on a single technology adoption framework to investigate cloud computing in PSOs, and eleven studies utilize combined, multi-framework approaches. Eight studies in the research corpus do not identify a specific theory. Tornatzky and Fleischer's technology, organization and environment framework (TOE) [53], either by itself or in combination with other theoretical approaches, emerges as the dominant framework for investigating organizational cloud computing adoption and use in public sector organizations (n=12). Rogers' diffusion of innovation (DOI) [54] (n=10), Davis' technology acceptance model (TAM) [55] (n=3), and institutional theory [56] (n=3) constitute other common theoretical frameworks in the body of research.

Geographically, all studies except for one relate to a single country as the focus of analysis; only [50] includes multiple countries in its analysis of public sector cloud adoption in five European nations. The research corpus spans across nineteen countries, with Australia (n=7), Saudi Arabia (n=4), and Iraq (n=3) emerging as the most-examined nations.

**TABLE 2: MAIN CHARACTERISTICS OF THE STUDIES INCLUDED IN THIS ANALYSIS**

| First Author | Country | Level of Government | Theory / Framework | Methodology | Data Collection Approach |
|---|---|---|---|---|---|
| Al Mudawi [24] | Saudi Arabia | Multiple | ACCE-GOV (merges TOE and DOI) | Quantitative | Survey with 838 Saudi government IT practitioners and government IT managers (Adoption of Cloud Computing in Saudi G-GOVernment) |
| Alsharari [25] | UAE | Central | ERP assimilation framework (institutional logics) | Qualitative | Case study of cloud enterprise resource planning system adoption in UAE state-owned investment development company, 15 semi-structure interviews |
| Mkhatshwa [26] | South Africa | Central and State | TOE | Qualitative | Survey and semi-structured interviews with 32 SA IT practitioners re cloud computing risks |
| Phuthong [27] | Thailand | Multiple | Factors derived from lit review | Quantitative | Survey questionnaire with 210 Thai government G-Cloud users |
| Ali [28] | Australia | Local | IS Complexity Framework | Mixed Method | 21 interviews with local gov IT managers and survey with 480 local government IT staff to investigate impact of complexity issues on cloud computing |
| Ali [29] | Australia | Local | TOE/Desires Framework/DOI | Quantitative | Survey with 480 local government IT staff to investigate key factors of cloud computing adoption |
| Ali [30] | Australia | Local | N/A | Mixed Method | 21 interviews with local gov IT managers and survey with 480 local government IT staff to investigate information security requirements affecting cloud computing adoption |
| Ali [31] | Australia | Local | N/A | Mixed Method | 21 interviews with local gov IT managers and survey with 480 local government IT staff to investigate critical factors surrounding the effect of government regulations on cloud adoption |
| Alzadjali [32] | Oman | Central | Institutional Theory | Qualitative | Case study utilizing 33 interviews with staff in Oman's Information Technology Authority and Ministry of Health to explore impact of institutional forces on G-Cloud adoption |
| Hashim [33] | Iraq | Central | TOE | Quantitative | Survey of 279 public university IT staff in Iraq to study critical factors affecting adoption. |
| Sallehudin [34] | Malaysia | Multiple | DOI/TOE/IS Success | Quantitative | Survey of 169 public sector IT practitioners to study cloud implementation |
| Alassafi [35] | Saudi Arabia | Multiple | N/A | Quantitative | Survey of 217 IT practitioners to validate security-related cloud adoption factors |
| Alkhlewi [36] | Saudi Arabia | Multiple | N/A | Mixed Method | Semi-structured interviews with 12 government IT experts and survey of 30 government IT practitioners in Saudi agencies |
| Branco [37] | Brazil | Central | N/A | Qualitative | Interviews with 12 government IT decision-makers from 3 public sector organizations to discern recommendations for successful cloud computing adoption. |
| Jones [38] | UK | Local | IS Benefit Model | Qualitative | Case study using participant observation and interviews in 3 local government PSOs |
| Mpanga [39] | Uganda | Local | Design Science Research | Qualitative | Focus groups with local government to develop factors for successful ERP implementation |
| Al-Ruithe [40] | Saudi Arabia | Multiple | N/A | Quantitative | Survey of 206 IT professionals in Saudi agencies to investigate cloud adoption concerns |
| Ali [41] | Australia | Local | TOE/Desires Framework/DOI | Qualitative | Interviews with 21 local government IT managers to determine cloud computing's potential for value creation |

| Khairuddin [42] | Malaysia | Central | TOE/DOI | Quantitative | Survey (structured questionnaire) of 93 government IT employees to confirm correlation of cloud adoption factors to actual adoption |
| Balasooriya [43] | Australia | Local | TOE/TAM | Quantitative | Survey of 200 IT professionals regarding factors influencing adoption |
| Shukur [44] | Iraq | Central | TOE/TAM | Quantitative | Survey of 25 IT professional in Iraqi government to study factors for cloud computing adoption |
| Liang [44] | China | State & Local | TOE | Qualitative | Semi-structured interviews with 24 government officials and IT managers and participant observation and review of government reports |
| Mohammed [45] | Yemen | Multiple | Fit Viability Model and DOI | Quantitative | Survey of 296 government IT managers to investigate factors that influence the decision to adopt in accordance with the fit viability model |
| Mohammed [46] | Yemen | Multiple | DOI and Task Technology Fit Model | Quantitative | Survey of 296 government IT managers to investigate factors that influence the decision to adopt in accordance with the fit viability model |
| Ali [47] | Australia | Local | N/A | Qualitative | Semi-structured interviews with 24 senior government IT managers |
| Maluleka [48] | South Africa | Central | DOI | Mixed Method | Survey of 28 government IT personnel and 6 interviews for triangulation. |
| El-Gazzar [49] | Norway | Multiple | Institutional Theory | Qualitative | Semi-structured interviews with 9 government IT staff and review of secondary data sources by the Norwegian government |
| Polyviou [50] | Five European Countries | Multiple | TOE/DOI | Qualitative | Semi-structured interviews with 21 government IT staff in five European countries |
| Wahsh [51] | Iraq | Central | TOE/DOI | Quantitative | Survey of 234 government IT personnel |
| Shin [52] | South Korea | Multiple | TAM | Mixed Method | Semi-structured interviews, focus groups, and survey of government IT personnel |

Structured along the lines of conventional public sector hierarchical levels derived from [57], [58], a majority of the studies in the body of research focus on a single level of government in their inquiry (n=17), with nine studies investigating local government organizations and eight studies examining central government organizations. Thirteen studies pursue a multi-level approach and investigate more than a single public sector level.

*C. Dominant Themes*

The dominant themes that emerged during the thematic synthesis of the included research studies are discussed below. These themes are responsive to the research question that guided this mixed methods analysis: What are the major themes, i.e. issues and factors, in the organizational adoption and use of cloud computing in public sector organizations? The themes were structured along technological, organizational, environmental, individual, and task-related focal points (*see* Fig. 2). In a divergence from prominent perspectives in the existing literature that traditionally conceive issues and factors either as barriers or enablers of cloud computing technology adoption [59]–[61], the themes identified below are conceptualized as flexible in terms of their effect. Depending on organizational management and execution, themes can present as enablers of or barriers to cloud computing adoption and use in PSOs in real world settings.

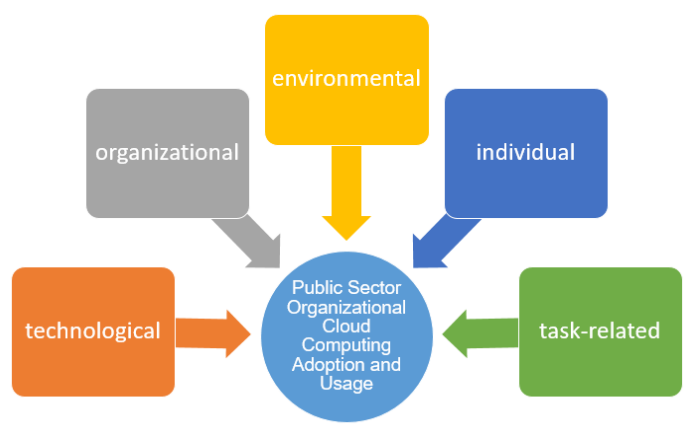

**Fig. 2: Thematic Structure of Issues and Factors that can facilitate or impede Cloud Computing Adoption in PSOs**

Table 3 summarizes the most prominent themes and subthemes that emerged during the thematic synthesis of the included research studies and lists items that occurred three or more times in the evidence base.

**TABLE 3: SUMMARY OF THEMES AND SUBTHEMES**

| Technological Theme | | Organizational Theme | | Environmental Theme | | Individual Theme | | Task-Related Theme | |
|---|---|---|---|---|---|---|---|---|---|
| Security | 23 | Leadership Support | 11 | Regulation/Policy | 17 | Trust | 6 | Alignment | 4 |
| Compatibility | 11 | Organizational Operations | 9 | Competitive Pressure | 5 | Ease of Use | 3 | | |
| Relative Advantage | 10 | Staff Skills | 6 | Standards | 5 | | | | |
| Cost | 9 | Project Management | 6 | Compliance | 3 | | | | |
| Infrastructure | 9 | Training | 3 | | | | | | |
| Complexity | 7 | Consultant Competence | 3 | | | | | | |
| Privacy | 6 | | | | | | | | |
| Data Sovereignty | 5 | | | | | | | | |
| Availability | 3 | | | | | | | | |

*D. Technological Themes*

The technological thematic focal point addresses the technological attributes of cloud computing that affect adoption and usage in public sector organizations and may also include the impact of other technologies that are in use in PSOs and that can affect the adoption and use of cloud computing [62]. Central themes in this focal point are security, compatibility, relative advantage and cost.

**Security** (n=23) emerges as the dominant theme in the technological thematic focal point. Unsurprisingly, as public sector organizations transition to and engage in cloud computing, they must exercise caution and ensure the safe and effective handling of government information. Cloud computing is inherently linked to security-relevant issues arising from system complexity, shared resource utilization, and change in control that—while conceptually not drastically different from security challenges in legacy computing environments—make security more challenging to achieve. The underlying systemic complexity of cloud computing that arises from the significant number of general computing and management components that are necessary to operate the environment enlarges the attack surface of the technology and increases vulnerability. The sharing of components and resources among different cloud tenants, while beneficial in terms of the economics of the technology, necessitates the expansive use of logical separation mechanisms that incur more vulnerabilities than physical separation in legacy environments. Additionally, with a transfer of responsibility to the cloud service provider, PSOs experience decreased control and visibility over the physical and logical features of the computing environment and become more reliant on the external service providers.

Despite these critical challenges, the security theme also encompasses potential security advantages for PSOs that might have positive effects on adoption and use. As the transition of security responsibilities to a cloud service provider as part of a shared security model frees up IT personnel resources in a PSO, the organization gains the opportunity to redirect its personnel towards high-risk areas that previously were underserved. The PSO is also likely to benefit from platform strength: as a result of economies of scale, cloud service providers are commonly able to provide more robust information assurance, security response, system management, and resiliency services compared to what PSOs who maintain their own in-house information technology are able to achieve.

**Compatibility** (n=11) is the second-most-dominant theme and focuses on the importance of determining how well cloud computing technology is compatible with the current IT infrastructure in PSOs and whether competing cloud computing services are interoperable with each other, allowing for future migrations of workloads from one cloud service provider to another and limiting the risk of vendor lock-in. The compatibility of information technology solutions has been a distinct issue in the study of technology adoption in the public sector as government organizations have been particularly challenged by an extensive footprint of obsolete systems and technology that is increasingly difficult to modernize and which commonly exceeds what can be found in the private sector, both in terms of quantity and complexity [63]. The more outdated and hence less compatible the legacy IT is with cloud computing, the more complex, costly, and risky does a migration to the cloud become. PSOs that have been able to keep their on-premise IT portfolios in a more current technological state face fewer compatibility issues, and PSOs that have taken the initiative to audit their IT assets have the ability and resources to cure existing technical debt, and understand their technology landscape can face compatibility as a strength that will benefit their adoption and use of cloud computing in line with its potential benefits.

The concerns regarding interoperability, i.e. the ability to transfer data or applications from one cloud platform to another, remains a substantial concern for PSOs and evolves around the ability to switch cloud service provides, utilize multiple service providers at the same time, link cloud services from different providers in a federated architecture, deploy hybrid cloud architectures, and expand cloud migration efforts. As cloud technologies mature further, competitive pressures and the evolution of standards and government-specific adoption frameworks will play a central role in advancing interoperability between different cloud service providers and critical government cloud applications.

**Relative Advantage** (n=10) emerges as the third central theme and deals with the extent to which cloud computing is viewed as being superior to legacy technology in terms of technological and operational advantages. With an enduring drumbeat that emphasizes the variety of benefits that cloud computing can confer, cloud service providers and public sector consultants have created the perception that cloud computing holds strong relative advantage for PSOs. The evidence base reflects this perception, with benefits like cost savings, operational improvements, and innovation capacity as prominent reasons for PSOs to pursue cloud computing technologies. While these benefits are possible and can in fact realize relative advantage, they are not produced automatically and require significant effort and planning to achieve. Frustration with the gaps between benefit expectations and actual migration outcomes is a sector-wide phenomenon [64].

**Cost** (n=9) constitutes another central theme and acknowledges the importance of addressing and understanding the financial implications of transitioning on-premise IT infrastructure to the cloud. While cost savings derived from the economies of scale that cloud computing offers are generally publicized as a key relative advantage of cloud computing and have been empirically validated in various settings [65]–[67], realizing and maintaining such cost savings requires disciplined planning and effective financial management.

The body of research highlights that cost-related challenges in public sector settings can inhibit cloud adoption and usage [38], [67] due to a failure by PSOs to accurately identify and quantify the costs associated with cloud computing. With the change from a capital expenditure financial model associated with legacy IT infrastructure to an operational expenditure financial model in the variable spend environment of cloud computing, PSOs need to re-architect key operational activities involving finance, planning, and acquisitions, to achieve effective financial control [68].

Additional, less prominent themes in the technological thematic focal point include **infrastructure** (n=9), **complexity** (n=7), **privacy** (n=6), **data sovereignty** (n=5) and **availability** (n=3). The infrastructure theme relates to the state of a PSO's IT infrastructure and the changes needed to fully support cloud computing operations, whereas the complexity theme focuses on the complexity of cloud computing technology use and adoption. Privacy focuses on privacy issues that derive from governmental mandates to safeguard personal privacy information and relates to the challenges of ensuring that the collection and storage of such information in cloud environments meets legal requirements. Data sovereignty relates to issues linked to the geographic location of cloud data collection and storage activities and the associated legal and regulatory requirements and limitations that arise as a result. The availability theme focuses on the potential impact of service and data availability in the cloud environment.

*E. Organizational Themes*

The organizational thematic focal point addresses salient organizational features of public sector organizations that affect cloud computing adoption and usage, capturing issues such as resource availability, organizational communications, and leadership behavior [62]. The central themes in this focal point are top leadership support and organizational operations.

**Leadership Support** (n=11) emerges as the dominant theme in the organizational thematic focal point, underscoring the instrumental role that PSO leaders play in the cloud adoption and usage process. PSO leaders direct, guide, and influence the organizational workforce towards the adoption and usage of cloud computing and are critical for ensuring the availability of necessary resources and services. PSO leaders are positioned to effect the essential redesign of organizational operations, creating the conducive operational environment required to support cloud computing in both the adoption and the operational stages. They also serve as communication conduits between external and internal stakeholders that influence the adoption process and are able to moderate competing views between these parties to minimize impacts on the execution of a cloud migration effort. While these leaders have the ability to create the momentum required for innovative technology change, their lack of support or actual resistance can on the other hand also hinder or interrupt adoption activities.

**Organizational Operations** (n=9) constitute the second central theme in this thematic focal point and address the link between operations factors and cloud adoption and use. PSO's internal operational capabilities and operational processes constitute an important element for a successful transition to and the effective and sustained use of cloud computing. Organizational processes and capabilities need to be able to support the financial operating model of cloud operations, reflect distinct legal and contracting requirements, and provide the technical and engineering competencies needed for a transition and sustainable use afterwards.

Additional, less prominent themes in the organizational thematic focal point include **staff skills** (n=6), **project management** (n=6), **training** (n=3), and **consultant competence** (n=3).

*F.  Environmental Themes*

This thematic focal point addresses the impact of a PSO's external environmental on cloud computing adoption and usage by public sector organizations, capturing salient issues such as the regulatory environment, technology trends in the public sector, and the existence of external standards [62].

**Regulation/Policy** (n=17) emerges as the dominant theme in this thematic focal point and addresses the impact of government policy—in the form of laws, regulations, and other governmental guidance—on the adoption and use of cloud computing. A robust legal, regulatory, and policy framework can resolve some of the risks and uncertainties that have hindered the governmental use of cloud solutions by providing clear guidance and enforceable requirements that address central issues such as information security and privacy, interoperability, liability, quality of services, and data sovereignty. Laws, regulations, and policies can also affect cloud adoption in a compulsory fashion if they set mandates such as the widely recognized "cloud first" or "cloud smart" policies.

Additional, less prominent themes in the environmental thematic focal point include **competitive pressure** (n=6), **standards** (n=5), and **compliance** (n=3). The competitive pressure theme relates to the impact that cloud adoption by public sector peers can have on PSOs in terms of not wanting to appear as laggards in the innovation sphere. The standards theme focuses on the importance of broadly adopted and sound technical standards for cloud computing that create a foundation for robust cloud computing migrations and operations without constraining continuous or future technology evolution. The compliance theme relates to the conformity of cloud solutions with the various legal and regulatory requirements that specifically apply to PSOs and that often impose more strenuous requirements compared to private sector cloud consumers, as well as the provision of tools by cloud service providers that ease how PSOs can demonstrate compliance to oversight authorities.

*G.  Individual Themes*

The individual thematic focal point addresses the attitudes and perceptions of key decision makers in public sector organizations that affect cloud computing adoption and usage even in organizational settings [69].

**Trust** (n=6) emerges as the dominant theme in this thematic focal point and relates to the confidence of key decision makers in PSOs that cloud computing, specific cloud computing solutions, or cloud service providers can deliver reliable, agile, responsive, and secure computing services that meet the requirements of the PSO. Without initial and continued trust, cloud computing adoption and sustained use is negatively affected. With trust, adoption and use are enabled. The trust theme evidences a mediating influence on PSO leadership support; when trust in cloud computing is not present, PSO leaders are unlikely to support or champion cloud computing initiatives.

An additional, less prominent theme in the individual thematic focal point is **ease of use** (n=3), relating to the perceptions of key PSO decision makers on whether cloud computing provides PSOs the functional utility needed to fulfill PSO missions while doing so with a low level of difficulty.

*H. Task-Related Themes*

The task-related thematic focal point focuses on the ability of cloud computing technology to meet work task demands and strategic operational demands that are present in public sector organizations and examines the impact that a good match between the technology and its' task applications and organizational strategy has on adoption and usage [70].

**Alignment** (n=4) emerges as the dominant and only significant theme in this thematic focal point and focuses on the importance of a continuous and dynamic alignment process between cloud computing and PSO strategy and operations, both prior to and after an adoption. Cloud computing, with its distinct potential benefits, cannot fully realize its potential without an adjustment of organizational tasks and strategy. A mismatch in the alignment of the technology can act as a barrier to adoption and further hinder assimilation in the post-adoption stage. In line with the flexible conceptualization of themes in this analysis, a well-matched alignment of cloud technology characteristics with organizational task applications and organizational strategy promotes adoption and use of the technology.

## IV.  CONCLUSION

This systematic, mixed methods analysis constitutes the initial phase of a larger research effort that involves forthcoming case studies of specific public sector cloud stakeholders; it identifies and synthesizes salient technological, organizational, environmental, individual, and task-related themes that affect cloud computing adoption and usage in a public sector context. As an analytical effort that is grounded in the peer-reviewed research base for public sector cloud computing, the salient themes span several theoretical technology adoption frameworks and acknowledge the importance of adopting diverse perspectives in order to capture the nuances and complexities of the technology in this distinct environment. Security, leadership support, regulation/policy, trust, and alignment emergence as the dominant issues and factors, restraining or enabling cloud computing adoption and use in these organizations depending on the organizational management and execution of the technology adoption process.

The technology and these salient, context-specific themes enable agencies to prioritize their adoption and use cases and highlight the importance of key issues. Government cloud practitioners can benefit from this analysis as it identifies the most salient themes from various perspectives and allows for an evidence-based assessment of which internal and external issues need to be considered during cloud adoption projects, providing a counter-balance to the industry-driven technology assertions that are marketed towards the public sector. For policymakers, this research reiterates the importance of the regulatory and policy environment for the expansion of cloud computing capabilities in the government sector and underscores the significance of robust cloud strategies that connect the successful execution of government functions with critical security, efficiency, and effectiveness factors.

Public sector organizations should focus their efforts on ensuring that they follow a risk-based strategy and expand their efforts to secure their cloud infrastructures. A defense-in-depth approach that pursues security in multiple layers and protects the confidentiality, integrity, and availability of government information as it travels across networks and resides on hardware is central to securing the cloud; additional steps such as the continuous monitoring of networks and the enhancement of information technology governance will further improve the security posture of government cloud initiatives.

With the importance of leadership support, the technical and non-technical leaders in public sector organizations are critical for pushing cloud adoption and use and underline the central role that the individuals in these leadership positions play. It is incumbent on government to attract, recruit, and retain leaders that comprehend technology and that have the leadership skills and abilities to direct, guide, and influence the organizational workforce towards the adoption and usage of cloud computing.

*A.  Limitations and Future Research*

The themes identified in this study present a snapshot in time and may shift as cloud computing technology further matures and as public sector adoptions and use continues to grow.

Due to the importance of cloud computing as an essential enabling technology for public sector organizations, it will undoubtedly be beneficial to continue the academic exploration of this topic and this present research effort constitutes one step in a larger research project that will further investigate cloud computing in a government context. As the empirical academic exploration of cloud computing technology adoption and use in the public sector is still lagging behind research in private sector settings, further academic attention is warranted and necessary.

# REFERENCES

[1]  P. Mell and T. Grace, "The NIST Definition of Cloud Computing," 2011. [Online]. Available: https://nvlpubs.nist.gov/nistpubs/Legacy/SP/nistspecialpublication800-145.pdf.

[2]  Insights, "Government Trends 2020 What are the most transformational trends in government today? A REPORT FROM THE DELOITTE CENTER FOR GOVERNMENT INSIGHTS About the Deloitte Center for Government Insights," 2020.

[3]  R. Agarwal, N. Khan, L. Santos, and G. Shenai, "How public-sector tech leaders can speed up the journey to the cloud," 2020. [Online]. Available: https://www.mckinsey.com/industries/public-and-social-sector/our-insights/how-public-sector-tech-leaders-can-speed-up-the-journey-to-the-cloud.

[4]  S. Marston, Z. Li, S. Bandyopadhyay, J. Zhang, and A. Ghalsasi, "Cloud computing - The business perspective," *Decis. Support Syst.*, vol. 51, no. 1, pp. 176–189, Apr. 2011, doi: 10.1016/j.dss.2010.12.006.

[5]  T. C. Chieu, A. Mohindra, A. A. Karve, and A. Segal, "Dynamic Scaling of Web Applications in a Virtualized Cloud Computing Environment," in *2009 IEEE International Conference on e-Business Engineering*, 2009, pp. 281–286, doi: 10.1109/ICEBE.2009.45.

[6]  "Building a culture of innovation to better serve citizens | AWS Public Sector Blog." https://aws.amazon.com/blogs/publicsector/building-culture-innovation-better-serve-citizens/ (accessed Apr. 20, 2021).

[7]  AWS, "Amazon Web Services: Risk and Compliance," 2020. [Online]. Available: https://d1.awsstatic.com/whitepapers/compliance/AWS_Risk_and_Compliance_Whitepaper.pdf.

[8]  M. Theby, "Public Sector Cloud Computing Adoption and Utilization During Covid-19: An Agenda for Research and Practice," *Int. J. Manag. Public Sect. Inf. Commun. Technol.*, vol. 13, no. 1, pp. 1–11, 2022, doi: 10.5121/ijmpict.2022.13101.

[9]  Shopp, "How the Pandemic Impacted Government's Cloud Migration Plans: The Good, the Bad, and the Ugly," 2021. https://www.nextgov.com/ideas/2021/06/how-pandemic-impacted-governments-cloud-migration-plans-good-bad-and-ugly/174602/ (accessed Aug. 01, 2021).

[10] N. Cannon, "Governments' Use of Cloud 2025: Fighting, Implementing or Optimizing?" Accessed: Jun. 26, 2022. [Online]. Available: https://www.gartner.com/8032bed4-0532-4042-a3e4-fdb65d178cbf.

[11] M. Theby, "Cloud Computing in the Public Sector: Mapping the Knowledge Domain," *Int. J. Manag. Public Sect. Inf. Commun. Technol.*, vol. 12, no. 4, pp. 1–17, 2021, doi: 10.5121/ijmpict.2021.12401.

[12] M. A. Wahsh and J. S. Dhillon, "A systematic review of factors affecting the adoption of cloud computing for e-Government implementation," *ARPN J. Eng. Appl. Sci.*, vol. 10, no. 23, pp. 17824–17832, 2015.

[13] A. Assaf, A. W. IisHamsir, and M. Muhammad, "Benefits and Risks of Cloud Computing in E-Government Tasks: A Systematic Review," *E3S Web Conf.*, vol. 328, p. 04005, 2021, doi: 10.1051/e3sconf/202132804005.

[14] Danielsen, L. S. Flak, and A. Ronzhyn, "Cloud Computing in eGovernment: Benefits and Challenges," *ICDS 2019 Thirteen. Int. Conf. Digit. Soc. eGovernments*, vol. 1, no. c, pp. 71–77, 2019.

[15] O. Abied, O. Ibrahim, and S. N.-I. Mat Kamal, "Adoption of Cloud Computing in E-Government: A Systematic Literature Review," *Pertanika J. Sci. Technol.*, vol. 30, no. 1, pp. 655–689, 2022, doi: 10.47836/pjst.30.1.36.

[16] M. Heyvaert, K. Hannes, and P. Onghena, *Using Mixed Methods Research Synthesis for Literature Reviews*. Thousand Oaks: Sage Publications, 2017.

[17] Aromataris and Z. Munn, *JBI MANUAL FOR EVIDENCE SYNTHESIS*. 2021.

[18] M. J. Page *et al.*, "The PRISMA 2020 statement: An updated guideline for reporting systematic reviews," *PLoS Med.*, vol. 18, no. 3, pp. 1–15, 2021, doi: 10.1371/JOURNAL.PMED.1003583.

[19] Q. N. Hong, A. Gonzalez-Reyes, and P. Pluye, "Improving the usefulness of a tool for appraising the quality of qualitative, quantitative and mixed methods studies, the Mixed Methods Appraisal Tool (MMAT)," *J. Eval. Clin. Pract.*, vol. 24, no. 3, pp. 459–467, Jun. 2018, doi: https://doi.org/10.1111/jep.12884.

[20] Q. N. Hong *et al.*, "Improving the content validity of the mixed methods appraisal tool: a modified e-Delphi study," *J. Clin. Epidemiol.*, vol. 111, pp. 49-59.e1, 2019, doi: 10.1016/j.jclinepi.2019.03.008.

[21] R. Skinner, R. Ryan Nelson, and W. W. Chin, "Synthesizing Qualitative Evidence: A Roadmap for Information Systems Research," *J. Assoc. Inf. Syst.*, vol. 23, no. 3, pp. 639–677, 2022, doi: 10.17705/1jais.00741.

[22] Thomas and A. Harden, "Methods for the thematic synthesis of qualitative research in systematic reviews," *BMC Med. Res. Methodol.*, vol. 8, pp. 1–10, 2008, doi: 10.1186/1471-2288-8-45.

[23] M. J. Page *et al.*, "The PRISMA 2020 statement: An updated guideline for reporting systematic reviews," *The BMJ*, vol. 372. BMJ Publishing Group, Mar. 29, 2021, doi: 10.1136/bmj.n71.

[24] N. Al Mudawi, N. Beloff, and M. White, "Developing a Framework of Critical Factors Affecting the Adoption of Cloud Computing in Government Systems (ACCE-GOV)," vol. 283. pp. 520–538, 2022, doi: 10.1007/978-3-030-80119-9_32.

[25] N. M. Alsharari, "Cloud computing and ERP assimilation in the public sector: institutional perspectives," *Transform. Gov. People, Process Policy*, vol. 16, no. 1, pp. 97–109, 2022, doi: 10.1108/TG-04-2021-0069.

[26] Mkhatshwa and T. Mawela, "Perceptions of Cloud Computing Risks in the Public Sector," vol. 419 LNNS. pp. 599–611, 2022, doi: 10.1007/978-3-030-96299-9_57.

[27] T. Phuthong, "Factors that influence cloud adoption in the public sector: The case of an emerging economy—Thailand," *Cogent Bus. Manag.*, vol. 9, no. 1, 2022, doi: 10.1080/23311975.2021.2020202.

[28] O. Ali, A. Shrestha, M. Ghasemaghaei, and G. Beydoun, "Assessment of Complexity in Cloud Computing Adoption: a Case Study of Local Governments in Australia," *Inf. Syst. Front.*, 2021, doi: 10.1007/s10796-021-10108-w.

[29] O. Ali, A. Shrestha, V. Osmanaj, and S. Muhammed, "Cloud computing technology adoption: an evaluation of key factors in local governments," *Inf. Technol. People*, vol. 34, no. 2, pp. 666–703, 2020, doi: 10.1108/ITP-03-2019-0119.

[30] O. Ali, A. Shrestha, A. Chatfield, and P. Murray, "Assessing information security risks in the cloud: A case study of Australian local government authorities," *Gov. Inf. Q.*, vol. 37, no. 1, 2020, doi: 10.1016/j.giq.2019.101419.

[31] O. Ali and V. Osmanaj, "The role of government regulations in the adoption of cloud computing: A case study of local government," *Comput. Law Secur. Rev.*, vol. 36, 2020, doi: 10.1016/j.clsr.2020.105396.

[32] Alzadjali and A. Elbanna, "Smart Institutional Intervention in the Adoption of Digital Infrastructure: The Case of Government Cloud Computing in Oman," *Inf. Syst. Front.*, vol. 22, no. 2, pp. 365–380, 2020, doi: 10.1007/s10796-019-09918-w.

[33] S. Hashim and Z. A. Al Sulami, "Cloud computing based e-government in Iraq using partial least square algorithm," *Indones. J. Electr. Eng. Comput. Sci.*, vol. 22, no. 2, pp. 345–352, 2020, doi: 10.11591/ijeecs.v22.i2.pp345-352.

[34] Sallehudin *et al.*, "Performance and key factors of cloud computing implementation in the public sector," *Int. J. Bus. Soc.*, vol. 21, no. 1, pp. 134–152, 2020, [Online]. Available: https://www.scopus.com/inward/record.uri?eid=2-s2.0-85085183394&partnerID=40&md5=052a15746599cc97d771ff396f917a0d.

[35] M. O. Alassafi, R. Alghamdi, A. Alshdadi, A. Al Abdulwahid, and S. T. Bakhsh, "Determining factors pertaining to cloud security adoption framework in government organizations: An exploratory study," *IEEE Access*, vol. 7, pp. 136822–136835, 2019, doi: 10.1109/ACCESS.2019.2942424.

[36] A. Alkhlewi, R. J. Walters, and G. Wills, "Towards a framework for the successful implementation of a government cloud in Saudi Arabia," *Int. J. Bus. Process Integr. Manag.*, vol. 9, no. 4, pp. 281–291, 2019, doi: 10.1504/IJBPIM.2019.105678.

[37] Branco T., I. Bianchi, and F. de Sá-Soares, "Cloud Computing Adoption in the Government Sector in Brazil: An Exploratory Study with Recommendations from IT Managers," vol. 11484 LNCS. pp. 162–175, 2019, doi: 10.1007/978-3-030-19223-5_12.

[38] S. Jones, Z. Irani, U. Sivarajah, and P. E. D. Love, "Risks and rewards of cloud computing in the UK public sector: A reflection on three Organisational case studies," *Inf. Syst. Front.*, vol. 21, no. 2, pp. 359–382, 2019, doi: 10.1007/s10796-017-9756-0.

[39] Mpanga and A. Elbanna, "A Framework for Cloud ERP System Implementation in Developing Countries: Learning from Lower Local Governments in Uganda," vol. 558. pp. 274–292, 2019, doi: 10.1007/978-3-030-20671-0_19.

[40] M. Al-Ruithe, E. Benkhelifa, and K. Hameed, "Key Issues for Embracing the Cloud Computing to Adopt a Digital Transformation: A study of Saudi Public Sector," in *Procedia Computer Science*, 2018, vol. 130, pp. 1037–1043, doi: 10.1016/j.procs.2018.04.145.

[41] O. Ali, J. Soar, and A. Shrestha, "Perceived potential for value creation from cloud computing: a study of the Australian regional government sector," *Behav. Inf. Technol.*, vol. 37, no. 12, pp. 1157–1176, 2018, doi: 10.1080/0144929X.2018.1488991.

[42] A. Khairuddin and A. F. Harun, "Cloud computing adoption in government agency," *Int. J. Eng. Technol.*, vol. 7, no. 3, pp. 157–162, 2018, doi: 10.14419/ijet.v7i3.15.17521.

[43] N. Prasanna Balasooriya, "A confirmatory investigation of the factors influencing the cloud adoption in local government organisations in Australia," in *ACIS 2018 - 29th Australasian Conference on Information Systems*, 2018, [Online]. Available: https://www.scopus.com/inward/record.uri?eid=2-s2.0-85071696225&partnerID=40&md5=ae2744b6beabb5490399b05570b4239e.

[44] B. S. Shukur, M. K. A. Ghani, and M. A. Burhanuddin, "An analysis of cloud computing adoption framework for Iraqi e-government," *Int. J. Adv. Comput. Sci. Appl.*, vol. 9, no. 8, pp. 104–112, 2018, doi: 10.14569/ijacsa.2018.090814.

[45] Mohammed, O. Ibrahim, M. Nilashi, and E. Alzurqa, "Cloud computing adoption model for e-government implementation," *Inf. Dev.*, vol. 33, no. 3, pp. 303–323, 2017, doi: 10.1177/0266666916656033.

[46] Mohammed, A. I. Alzahrani, O. Alfarraj, and O. Ibrahim, "Cloud Computing Fitness for E-Government Implementation: Importance-Performance Analysis," *IEEE Access*, vol. 6, pp. 1236–1248, 2017, doi: 10.1109/ACCESS.2017.2778093.

[47] O. Ali, J. Soar, and J. Yong, "An investigation of the challenges and issues influencing the adoption of cloud computing in Australian regional municipal governments," *J. Inf. Secur. Appl.*, vol. 27–28, pp. 19–34, 2016, doi: 10.1016/j.jisa.2015.11.006.

[48] S. M. Maluleka and N. Ruxwana, "Cloud computing as an alternative solution for South African public sector: A case for department of social development," vol. 444. pp. 481–491, 2016, doi: 10.1007/978-3-319-31232-3_45.

[49] R. F. El-Gazzar and F. Wahid, "Strategies for cloud computing adoption: Insights from the Norwegian public sector," in *Proceedings of the 12th European, Mediterranean and Middle Eastern Conference on Information Systems, EMCIS 2015*, [Online]. Available: https://www.scopus.com/inward/record.uri?eid=2-s2.0-85084018175&partnerID=40&md5=915b1ad3cade9a5572d9d1c4d89a2bd3.

[50] A. Polyviou and N. Pouloudi, "Understanding cloud adoption decisions in the public sector," in *Proceedings of the Annual Hawaii International Conference on System Sciences*, 2015, vol. 2015-March, pp. 2085–2094, doi: 10.1109/HICSS.2015.250.

[51] A. Wahsh and J. S. Dhillon, "An investigation of factors affecting the adoption of cloud computing for E-government implementation," in *2015 IEEE Student Conference on Research and Development, SCOReD 2015*, pp. 323–328, doi: 10.1109/SCORED.2015.7449349.

[52] D. H. Shin, "User centric cloud service model in public sectors: Policy implications of cloud services," *Gov. Inf. Q.*, vol. 30, no. 2, pp. 194–203, 2013, doi: 10.1016/j.giq.2012.06.012.

[53] Tornatzky and M. Fleischer, *The Process of Technology Innovation*. Lexington, MA: Lexington Books, 1990.

[54] E. M. Rogers, *Diffusion of Innovations*, 5th ed. New York, NY: Free Press, 2003.

[55] Davis, "Perceived usefulness, perceived ease of us, and user acceptance of information technology," *MIS Q.*, vol. 13, pp. 319–340, 1989, doi: https://doi.org/10.2307/249008.

[56] Meyer and B. Rowan, "Institutionalized organizations: Formal structure as myth and ceremony.," *Am. J. Sociol.*, vol. 83, pp. 340–363, 1977, doi: http://dx.doi.org/10.1086/226550.

[57] OECD, *Government at a Glance 2021*. Paris, France: OECD Publishing, 2021.

[58] IMF, *Government Finance Statistics Manual 2014*. Washington, DC: International Monetary Fund, 2014.

[59] Metheny, *Federal cloud computing: The definitive guide for cloud service providers*. Amsterdam: Syngress, 2017.

[60] V. Kumar and R. Vidhyalakshmi, "Cloud computing," in *Reliability aspect of cloud computing environment*, V. Kumar and R. Vidhyalakshmi, Eds. Singapore: Springer Singapore, 2018, pp. 1–28.

[61] T. Ananth Kumar, T. Arun Samuel, R. Jackson Samuel, and M. Niranjanamurthy, Eds., *Privacy and security challenges in cloud computing: A holistic approach*. Boca Raton: CRC Press, 2022.

[62] J. Baker, "The technology–organization–environment framework," in *Information systems theory: Explaining and predicting our digital society*, Y. K. Dwivedi, M. R. Wade, and S. L. Schneberger, Eds. New York, NY: Springer New York, 2012, pp. 231–245.

[63] United States Government Accountability Office, "Agencies Need to Develop and Implement Modernization Plans for Critical Legacy Systems," Washington, DC, 2019. [Online]. Available: https://www.gao.gov/assets/gao-19-471.pdf.

[64] J. Taillon, J. Mariani, and D. Bourgeois, "Government cloud: A mission accelerator for future innovation," Washington, DC, 2019. [Online]. Available: https://www2.deloitte.com/us/en/insights/industry/public-sector/government-cloud-innovation.html.

[65] Ali, A. Shrestha, V. Osmanaj, and S. Muhammed, "Cloud computing technology adoption: an evaluation of key factors in local governments," *Inf. Technol. People*, 2020, doi: 10.1108/ITP-03-2019-0119.

[66] S. Jones, Z. Irani, U. Sivarajah, and P. E. D. Love, "Risks and rewards of cloud computing in the UK public sector: A reflection on three Organisational case studies," *Inf. Syst. Front.*, vol. 21, no. 2, pp. 359–382, 2019, doi: 10.1007/s10796-017-9756-0.

[67] Y. Liang, G. Qi, K. Wei, and J. Chen, "Exploring the determinant and influence mechanism of e-Government cloud adoption in government agencies in China," *Gov. Inf. Q.*, vol. 34, no. 3, pp. 481–495, 2017, doi: 10.1016/j.giq.2017.06.002.

[68] Theby, "Public Sector Cloud Computing Adoption and Utilization During Covid-19: An Agenda for Research and Practice," *Int. J. Manag. Public Sect. Inf. Commun. Technol.*, vol. 13, no. 1, pp. 1–11, Mar. 2022, doi: 10.5121/ijmpict.2022.13101.

[69] D. Williams, N. P. Rana, and Y. K. Dwivedi, "A bibliometric analysis of articles citing the unified theory of acceptance and use of technology," in *Information systems theory: Explaining and predicting our digital society*, Y. K. Dwivedi, M. R. Wade, and S. L. Schneberger, Eds. New York, NY: Springer New York, 2012, pp. 37–62.

[70] B. Furneaux, "Task-technology fit theory: A survey and synopsis of the literature," in *Information systems theory: Explaining and predicting our digital society*, Y. K. Dwivedi, M. R. Wade, and S. L. Schneberger, Eds. New York, NY: Springer New York, 2012, pp. 87–106.